\documentclass[a4paper]{jpconf}
\usepackage{graphicx}
\usepackage{amsmath}
\usepackage[colorlinks=true, allcolors=blue]{hyperref}
\begin{document}
\title{Spin polarization of $\Lambda$ hyperons in $e^+e^-\to\Lambda\bar{\Lambda}$ at BESIII}

\author{Cui Li$^1$ and Karin Sch\"{o}nning for the BESIII Collaboration}

\address{Department of Physics and Astronomy, Uppsala University, Box 516, 75120 Uppsala, Sweden}

\ead{$^1$cui.li@physics.uu.se} \ead{$^2$karin.schonning@physics.uu.se}
\begin{abstract}
The BESIII collaboration here reports the first observation of polarized $\Lambda$ and $\bar{\Lambda}$ hyperons produced in two different processes: i) the resonant $e^+e^- \to J/\psi\to\Lambda\bar{\Lambda}$, using a data sample of 1.31 $\times$ 10$^9$ $J/\psi$ events and ii) the non-resonant $e^+e^-\to \gamma^* \to \Lambda\bar{\Lambda}$, using a 66.9 pb$^{-1}$ data sample collected at $\sqrt{s} =$ 2.396 GeV. In $e^+e^-\to J/\psi\to\Lambda\bar{\Lambda}$, the phase between the electric and the magnetic amplitude is measured for the first time to be $42.3^{\mathrm{o}}\pm0.6^{\mathrm{o}}\pm0.5^{\mathrm{o}}$. The multi-dimensional analysis enables a model-independent measurement of the decay parameters for $\Lambda\to p\pi^-$ ($\alpha_-$), $\bar{\Lambda}\to\bar{p}\pi^+$ ($\alpha_+$) and $\bar{\Lambda}\to\bar{n}\pi^0$ ($\bar{\alpha}_0$). The obtained value $\alpha_-=0.750\pm0.009\pm0.004$ differs with ~5$\sigma$ from the PDG value. This value, together with the measurement $\alpha_+=-0.758\pm0.010\pm0.007$ allow for the most precise test of CP violation in $\Lambda$ decays so far: $A_{CP}=(\alpha_-+\alpha_+)/(\alpha_--\alpha_+)$ of $-0.006\pm0.012\pm0.007$. The decay asymmetry $\bar{\alpha}_0=-0.692\pm0.016\pm0.006$ is measured for the first time. The $e^+e^-\to\Lambda\bar{\Lambda}$ reaction at $\sqrt{s} =$ 2.396 GeV enables a first complete measurement of the time-like electric and magnetic form factor of any baryon, of the modulus of the ratio $R=|G_E/G_M|$ and of the relative phase $\Delta\Phi=\Phi_E-\Phi_M$. With the decay asymmetry parameters from the $J/\psi$ data, the obtained values are $R=0.96\pm0.14\pm0.02$ and $\Delta\Phi=37^{\mathrm{o}}\pm12^{\mathrm{o}}\pm6^{\mathrm{o}}$. In addition, the cross section has been measured with unprecedented precision to be $\sigma = 119.0\pm 5.3\pm5.1$~pb, which corresponds to an effective form factor of $|G|=0.123\pm0.003\pm0.003$.  

\end{abstract}

\section{Introduction}

Many of the most challenging questions in contemporary physics manifest themselves in our lack of understanding of one of the most abundant building blocks of the Universe: the nucleon. Despite being known for a century, we still do not understand its abundance~\cite{sakharov}, its mass~\cite{pmass}, its spin~\cite{pspin}, its size~\cite{pradius} nor its intrinsic structure~\cite{pstructure}. In this work, we use hyperons as a diagnostic tool to shed light on two of these issues: nucleon abundance and nucleon structure. The idea is to learn more about a system by making a small change to it and see how the system reacts. More specifically we replace one of the building light quarks in the nucleon, by the heavier strange quark. Let's first consider the \textit{nucleon abundance}, more commonly known as the matter-antimatter asymmetry of the Universe. Assuming equal amounts being created in the Big Bang, the amount of matter (\textit{e.g.} nucleons) has been enriched with respect to anti-matter (antinucleons). One of three necessary criteria for this is the existence of CP violating processes. So far, CP violation beyond the Standard Model has never been observed for baryons. \textit{Nucleon structure} has been studied since the 1960's~\cite{ffreview} and one very intriguing finding is the charge distribution of the neutron: it is negative in the center and at the rim, positive in between~\cite{pstructure}. For a ground-state wave function this amount of wiggles is surprising. It is desirable to measure charge distributions also for hyperons, to test SU(3) flavor symmetry and diquark correlations~\cite{jaffe}. Hadron structure is parameterized in terms of \textit{form factors}. In the case of nucleons, \textit{space-like} form factors can be studied in elastic scattering. For the unstable hyperons, this is experimentally unfeasible, but instead, \textit{time-like} form factors can be studied in $e^+e^-$ annihilations. Space-like and time-like form factors are connected by dispersion relations.

The production of spin 1/2 baryon-antibaryon pairs from $e^+e^-$ annihilation with an intermediate vector (\textit{e.g.} $\gamma^*$ or $J/\psi$) follow the same formalism since there are two contributing production amplitudes. In the case of $J/\psi$, we refer to these amplitudes as electric and magnetic hadron amplitudes and in the case of $\gamma^*$ as the electric and magnetic form factors (EMFFs) $G_E(q^2)$ and $G_M(q^2)$. In the time-like region, where the square of the four-momentum transfer $s=q^2$ is positive, the amplitudes can be complex with a relative phase $\Delta\Phi=\Phi_E-\Phi_M$. A non-zero relative phase between the parameters implies spin polarization of the baryons even if the colliding beams are unpolarized. Hyperons provide a unique opportunity since their polarization, in contrast to nucleons, is experimentally accessible thanks to their weak, parity violating and thereby self-analyzing decays. In the case of $\Lambda\to p\pi^-$ decays, the angular distribution of the daughter protons is given by $1+\alpha_-\mathbf{P}_{\Lambda}\cdot\hat{\mathbf{n}}$, where $\hat{\mathbf{n}}$ is a unit vector along the proton momentum in the $\Lambda$ rest frame, $\mathbf{P}_{\Lambda}$ the $\Lambda$ hyperon polarization vector and $\alpha_-$ is the decay asymmetry parameter~\cite{TDLee}. The decay parameters $\alpha_+$ for $\bar{\Lambda}\to\bar{p}\pi^+$, $\alpha_0$ for $\Lambda\to n\pi^0$, and $\bar{\alpha}_0$ for $\bar{\Lambda}\to\bar{n}\pi^0$~\cite{PDG} are defined in the same way. The detailed formalism for $e^+e^-$ production, with the subsequent decay of the hyperon into a baryon and a pseudoscalar meson, has been outlined in Ref.~\cite{Faldt:2016qee}. In short, the joint angular distribution of the decay chain $e^+e^-\to(J/\psi)/\gamma^* \to\Lambda\bar{\Lambda}$ ($\Lambda\to p\pi^-$ and $\bar{\Lambda}\to\bar{p}\pi^+$) can be expressed in terms of $\Delta\Phi$ and the angular distribution parameter $\eta$: 
\begin{equation} 
\label{angdist}
\begin{split}
{\cal{W}}({\boldsymbol{\xi}})&={\cal{T}}_0+{{{\eta}}}{\cal{T}}_5\\
&+{{\alpha_-\alpha_+}}\left({\cal{T}}_1
+\sqrt{1-{{\eta}}^2}\cos({{\Delta\Phi}}){\cal{T}}_2
+{{\eta}}{\cal{T}}_6\right)\\
&+\sqrt{1-{{\eta}}^2}\sin({{\Delta\Phi}})
\left(\alpha_-{\cal{T}}_3+\alpha_+{\cal{T}}_4\right).
\end{split}
\end{equation}

Here, $\boldsymbol{\xi}$ represents the five measured quantities $\theta$ ($\Lambda$ scattering angle), $\theta_1$ and $\phi_1$ (decay proton angles) and $\theta_2$ and $\phi_2$ (decay antiproton angles). Details of the reference system are given in Ref. \cite{faldtkupsc}. The seven functions ${\cal{T}}_k({\boldsymbol{\xi}})$ do not depend on the physical quantities $\eta$ and $\Delta\Phi$ but only on the measured angles:
\begin{align}
	{\cal{T}}_0({\boldsymbol{\xi}}) =&1,\nonumber\\
	{\cal{T}}_1({\boldsymbol{\xi}}) =&{\sin^2\!\theta}\sin\theta_1\sin\theta_2\cos\phi_1\cos\phi_2+
{\cos^2\!\theta}\cos\theta_1\cos\theta_2,\nonumber\\
	{\cal{T}}_2({\boldsymbol{\xi}}) =&{\sin\theta\cos\theta}\left(\sin\theta_1\cos\theta_2\cos\phi_1+
\cos\theta_1\sin\theta_2\cos\phi_2\right),\nonumber\\
	{\cal{T}}_3({\boldsymbol{\xi}}) =&{\sin\theta\cos\theta}\sin\theta_1\sin\phi_1,\nonumber\\
	{\cal{T}}_4({\boldsymbol{\xi}}) =&{\sin\theta\cos\theta}\sin\theta_2\sin\phi_2,\nonumber\\
    {\cal{T}}_5({\boldsymbol{\xi}}) =&{\cos^2\!\theta},\nonumber\\
	{\cal{T}}_6({\boldsymbol{\xi}}) =&{\cos\theta_1\cos\theta_2
    -\sin^2\theta\sin\theta_1\sin\theta_2\sin\phi_1\sin\phi_2}.\nonumber
\end{align}
This means that Eq. \ref{angdist} has one term (${\cal{T}}_0 + \eta {\cal{T}}_5$) that only depends on the $\Lambda$ scattering angle, one term ($\sqrt{1-{{\eta}}^2}\sin({{\Delta\Phi}})(\alpha_-{\cal{T}}_3+\alpha_+{\cal{T}}_4)$) that describes the transverse polarization and one term ($\alpha_-\alpha_+({\cal{T}}_1+\sqrt{1-{{\eta}}^2}\cos({{\Delta\Phi}} ){\cal{T}}_2+{{\eta}}{\cal{T}}_6)$) that describes the spin correlation of the produced hyperons. In the case of $J/\psi\to\Lambda\bar{\Lambda}$, the value of $\eta$ has been measured with high precision {\it e.g.} by the BES~\cite{alphapsibes}, BESII~\cite{alphapsibes2} and BESIII Collaborations~\cite{alphapsibes3}. In the case of non-resonant $e^+e^-\to\Lambda\bar{\Lambda}$ production, $\eta$ is related to the form factor ratio $R=|G_E|/|G_M|$ that has been measured with large uncertainties by the BaBar collaboration~\cite{BaBarll}. However, so far there were no attempts to extract the phase $\Delta\Phi$ in any of the cases.

In this report, we present the first observation of the spin polarization of $\Lambda$ and $\bar{\Lambda}$ hyperons from the $J/\psi\to\Lambda\bar{\Lambda}$ decay, using a data sample of $(1310.6\pm7.0)\times10^6 J/\psi$ events~\cite{jpsinumber}. The polarized hyperons are used to determine decay parameters of $\Lambda$ and $\bar{\Lambda}$.  We also present the first complete determination of the time-like $\Lambda$ EMFFs using a 66.9 pb$^{-1}$ sample collected at $\sqrt{s}$ = 2.396 GeV. Both data samples are collected by the BESIII collaboration.

\section{The BESIII Experiment}
The Beijing Spectrometer III (BESIII) is an integrated part of the Beijing Electron Positron Collider II (BEPCII)~\cite{bes}, a double ring $e^{+}e^{-}$ collider operating at $2.0-4.6$ GeV c.m. energies with a design luminosity of $1 \times 10^{33}$ cm$^{-2}\textrm{s}^{-1}$ at c.m. energy of 3.773 GeV. BESIII is a $4\pi$ detector that has accumulated world-leading samples of $J/\psi$, $\psi(2S)$, $\psi(3770)$ events for the study of light hadron and charmonium spectroscopy. Furthermore, BESIII has collected unprecedented samples of non-resonant data for a broad scope of studies of low-energy QCD, in particular hadron EMFFs and charmonium-like states \textit{e.g.} $X$, $Y$ and $Z$~\cite{besphysics}. 

\section{Decay parameters of $\Lambda$ hyperons}

\begin{figure}[h]
\centering
\includegraphics[width=16cm]{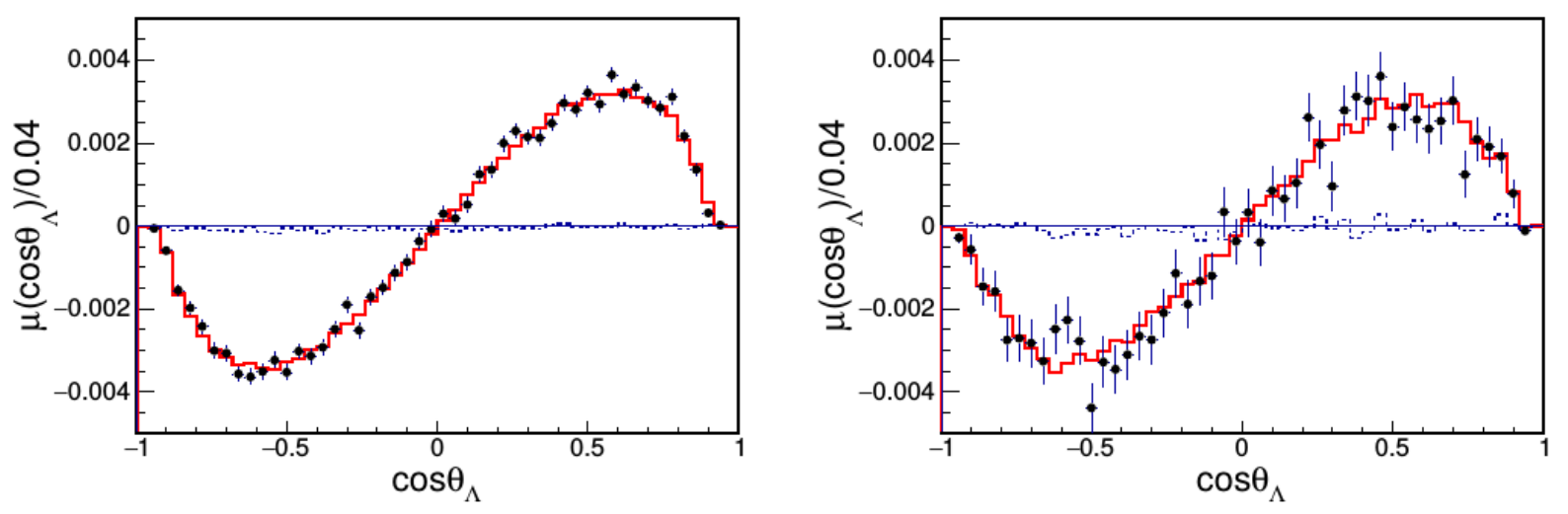}
\put(-380,120){(a)}
\put(-380,100){Preliminary}
\put(-160,120){(b)}
\put(-160,100){Preliminary}
\caption{Moments $\mu(\cos\theta_{\Lambda})$ as a function of $\cos\theta_{\Lambda}$ for (a) $p\pi^-\bar{p}\pi^+$ and (b) $p\pi^-\bar{n}\pi^0$ data sets. The data are not corrected for acceptance. The filled dots represent data and the solid-line histogram the global fit result. The dotted histogram represent unpolarized MC events.} 
\label{jpsiPol} 
\end{figure}
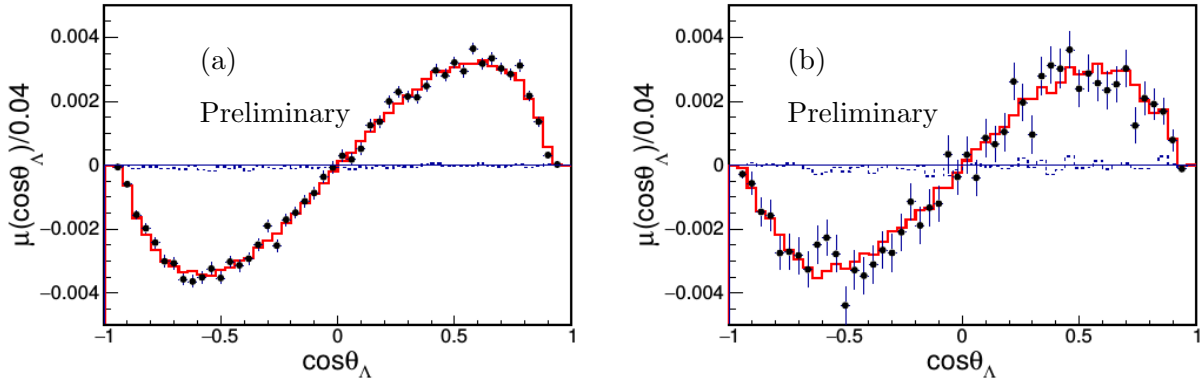

In this study, events from the $e^+e^- \to J/\psi\to(\Lambda\to p\pi^-)(\bar{\Lambda}\to\bar{p}\pi^+)$ and $e^+e^- \to J/\psi\to(\Lambda\to p\pi^-)(\bar{\Lambda}\to\bar{n}\pi^0)$ reactions are identified and analyzed. The joint angular distribution for $J/\psi\to(\Lambda\to p\pi^-)(\bar{\Lambda}\to\bar{n}\pi^0)$ is obtained by replacing $\alpha_+$ by $\bar{\alpha}_0$ in Eq.~\ref{angdist}. A simultaneous log-likelihood fit is performed to the final data samples, comprising 420593 and 47009 events with estimated background of 399$\pm$20 and 66$\pm$8 events, respectively. The likelihood function is constructed from the product of probability density functions, that for the $i$:th event is defined by $P({\boldsymbol{\xi_i}})$ = ${\cal{C}}(\eta, \Delta\Phi, \alpha_-, \alpha_2){\cal{W}}({\boldsymbol{\xi_i}})$. Here, $\alpha_2 = \alpha_+$ and $\alpha_2 = \bar{\alpha}_0$ for $p\pi^-\bar{p}\pi^+$ and $p\pi^-\bar{n}\pi^0$, respectively. The measured quantities in the $i$:th event are represented by the vector ${\boldsymbol{\xi_i}}$ and ${\cal{W}}({\boldsymbol{\xi_i}})$ is given by Eq.~\ref{angdist}. The normalization factor ${\cal{C}}(\eta, \Delta\Phi, \alpha_-, \alpha_2)$ is calculated for each parameter set using as a sum of the corresponding ${\cal{W}}({\boldsymbol{\xi}})$ for events generated according to phase space, propagated through the virtual detector and selected using the same criteria as the real data. The main sources of systematic uncertainties are tracking, $\Lambda(\bar{\Lambda})$ and $\pi^0$ reconstruction efficiencies, the kinematic fit and in the $\bar{\Lambda}\to\bar{n}\pi^0$ case, the identification of the $\bar{n}$ shower. More details can be found in Ref. \cite{llbararxiv}.
A spin polarization of $\Lambda/\bar{\Lambda}$ is observed with the dependence of the scattering angle from Eq. \ref{angdist} and from this, the phase is determined to be $\Delta\Phi=(42.4\pm0.6\pm0.5)^{\mathrm{o}}$. The polarization is illustrated in Fig.~\ref{jpsiPol} that shows the moment $\mu(\cos\theta_{\Lambda})=1/N\Sigma^{N(\theta_{\Lambda})}_i(\sin\theta^i_1\sin\phi^i_1-\sin\theta^i_2\sin\phi^i_2)$ as a function of the scattering angle $\cos\theta_{\Lambda}$. Here, $N$ is the total number of events in the data sample and $N(\theta_{\Lambda})$ is the number of events in a $\cos\theta_{\Lambda}$ bin.

The results for $\eta, \alpha_-, \alpha_+$ and $\bar{\alpha}_0$ are given in Table~\ref{jpsiresult}. The most striking part is that $\alpha_-$ = 0.750$\pm$0.009$\pm$0.004, \textit{i.e.} $\sim5\sigma$ larger than the PDG value. The latter is based on measurements from the 1960's and 1970's using a method of decay proton polarization and with data samples of $\approx$10000 \cite{PDG}. Our value is statistically more precise and obtained using a model-independent approach. The ratio $\bar{\alpha}_0/\alpha_+$ = 0.913$\pm$0.028$\pm$0.012 obtained from our data is 3$\sigma$ lower than +1.00 as predicted by the $|\Delta I|$ = $\frac{1}{2}$ rule for non-leptonic decays. This deviation is larger than expected due to radiative corrections~\cite{DeltaI1,DeltaI2}. Finally, from our decay parameters, the so far most precise CP test for $\Lambda$ decays is performed. The $CP$ odd observable $A_{CP} = (\alpha_-+\alpha_+)/(\alpha_--\alpha_+)$ was calculated and found consistent with zero as shown in the Table~\ref{jpsiresult}.
\begin{table}[h]
\caption{\label{jpsiresult}Summary of the results: the $J/\psi\to\Lambda\bar{\Lambda}$ angular distribution parameter $\eta$, the phase $\Delta\Phi$, the asymmetry parameters for the $\Lambda\to p\pi^-, \bar{\Lambda}\to\bar{p}\pi^+$ and $\bar{\Lambda}\to\bar{n}\pi^0$ decays, the $CP$ asymmetry $A_{CP}$, and the ratio $\bar{\alpha}_0/\alpha_+$. The first uncertainty is statistical, and the second one is systematic.}
\begin{center}
\begin{tabular}{c c c}
\br
Parameters & Preliminary results & Previous results\\
$\eta$&0.461$\pm$0.006$\pm$0.007&0.469$\pm$0.027~\cite{alphapsibes3}\\
$\Delta\Phi$(rad)&0.740$\pm$0.010$\pm$0.008&$-$\\
\mr
$\alpha_-$&0.750$\pm$0.009$\pm$0.004&0.642$\pm$0.013~\cite{PDG}\\
$\alpha_+$&-0.758$\pm$0.010$\pm$0.007&-0.71$\pm$0.08~~~\cite{PDG}\\
$\bar{\alpha}_0$&-0.692$\pm$0.016$\pm$0.006&$-$\\
$A_{CP}$&-0.006$\pm$0.012$\pm$0.007&0.006$\pm$0.021~\cite{PDG}\\
$\bar{\alpha}_0/\alpha_+$&0.913$\pm$0.028$\pm$0.012&$-$\\
\br
\end{tabular}
\end{center}
\end{table}

\section{Time-like EMFFs of $\Lambda$ hyperons}

The EMFFs are measured by selecting and analyzing $e^+e^-\to\Lambda\bar{\Lambda}, \Lambda\to p\pi^-, \bar{\Lambda}\to\bar{p}\pi^+$ events at a c.m. energy of $\sqrt{s}$ = 2.396 GeV. The parameters $\eta$ and $\Delta\Phi$ are extracted using the same method as in the $J/\psi \to \Lambda \bar{\Lambda}$ but keeping the decay parameters fixed, first to $\alpha_-=-\alpha_+=0.642$, then to $\alpha_-=-\alpha_+=0.750$
The form factor ratio $R=|G_E/G_M|$ is related to the scattering angle parameter $\eta$ by $\eta=\frac{\tau-R^2}{\tau+R^2}$. 

\begin{figure*}[!htbp]
\begin{center}
\includegraphics[width=3.0in,height=2.22in,angle=0]{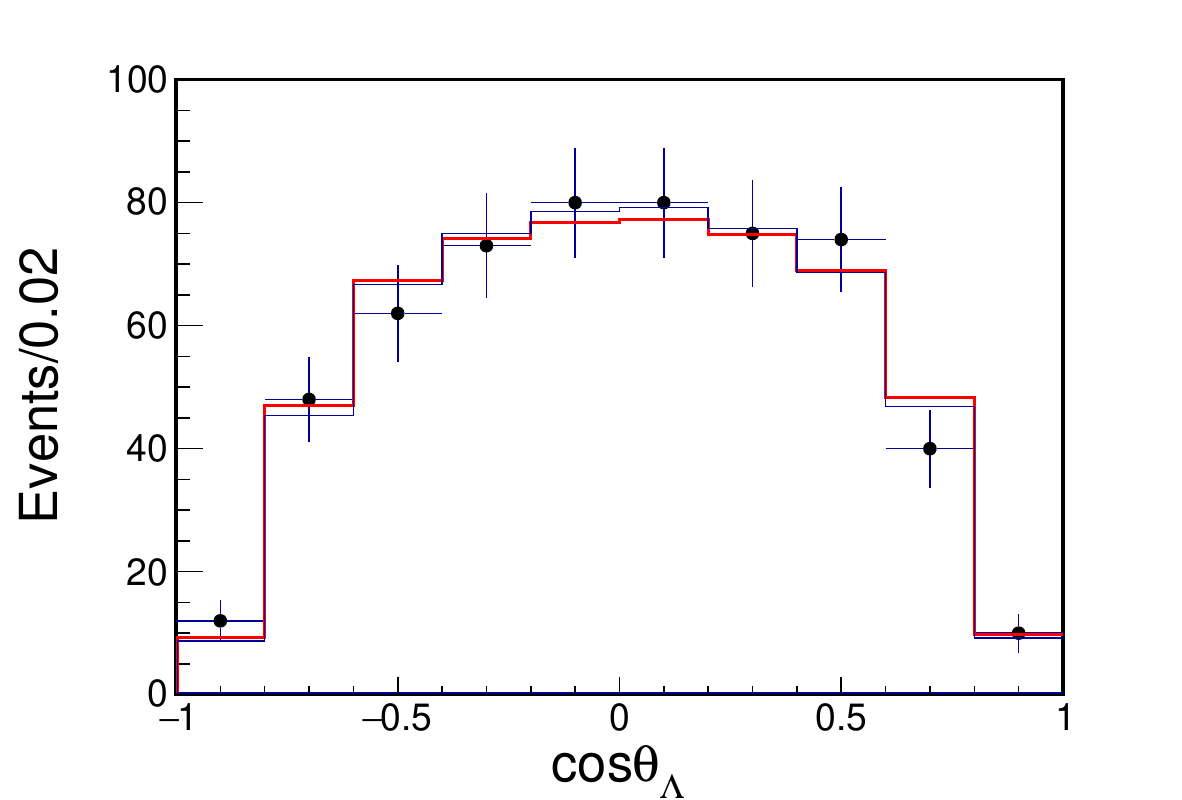}
\put(-50, 120){(a)}
\put(-130, 40){Preliminary}
\includegraphics[width=3.0in,height=2.22in,angle=0]{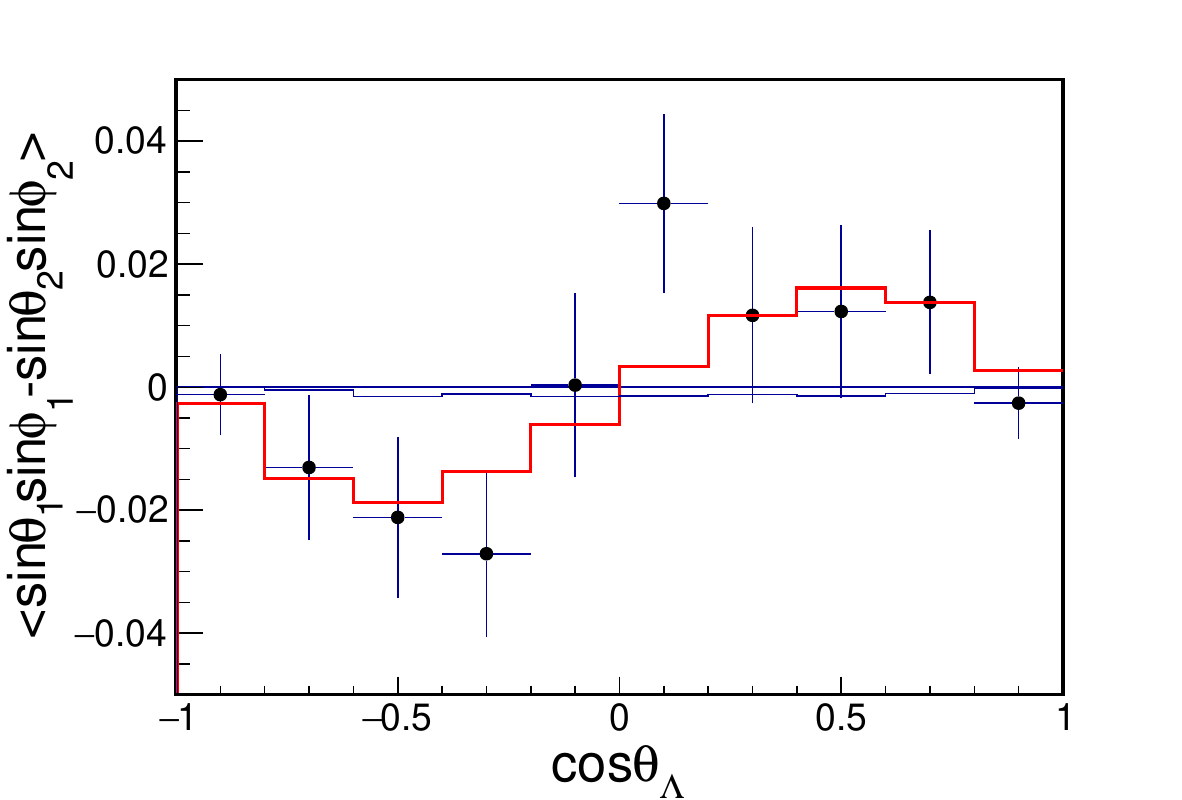}
\put(-50, 120){(b)}
\put(-120, 40){Preliminary}
\caption{The $\Lambda$ angular distribution (a) and the polarization effect as a function of $\cos\theta_{\Lambda}$ (b).}
\label{fig:e2396Pol}
\end{center}
\vspace{-5mm}
\end{figure*}

A scattering angle dependent spin polarization of $\Lambda/\bar{\Lambda}$ is observed as illustrated in Fig.~\ref{fig:e2396Pol}(b), implying a non-zero phase. Using $\alpha_-=0.642$~\cite{PDG}, we obtain $R=0.94\pm0.16\pm0.03$ and $\Delta\Phi=42^{\mathrm{o}}\pm16^{\mathrm{o}}\pm8^{\mathrm{o}}$ from the log-likelihood fit. With $\alpha_-=-\alpha_+=0.750$ the results become $R=0.96\pm0.14\pm0.02$ and $\Delta\Phi=37^{\mathrm{o}}\pm12^{\mathrm{o}}\pm6^{\mathrm{o}}$. The main sources of systematics are found to be the kinematic fit, the invariant mass cut and the scattering angle range of the fit. 


The Born cross section, that assumes one-photon exchange, is straight-forwardly parameterized in terms of $G_E$ and $G_M$:
\begin{equation}
\sigma_{B\bar{B}}(q^2) = \frac{4\pi\alpha^2\beta}{3s}[|G_M(q^2)|^2+\frac{1}{2\tau}|G_E(q^2)|^{2}]
\label{equ-borncs}
\end{equation}
where $\alpha$=1/137.036 is the fine-structure constant, $\beta=\sqrt{1-4m^2_Bc^4/s}$ the velocity, \textit{c} the speed of light, \textit{s} the square of the center-of-mass (c.m.) energy, $m_B$ the mass of the baryon and $\tau = s/(4m^2_B)$. The Coulomb correction factor \textit{C}~\cite{C1}, accounts for the final state electromagnetic interaction of charged point-like fermion pairs and neutral baryons. Experimentally, the cross section is extracted using $\sigma_{\Lambda\bar{\Lambda}} = \frac{N_{signal}}{\mathcal{L}_{int}\epsilon(1+\delta)\mathcal{B}}$, where $N_{signal}=N_{data}-N_{bg}$, $N_{data}$ is the number of selected events, $N_{bg}$ is the background estimated from sidebands, $\mathcal{L}_{int}$ is the integrated luminosity, $\epsilon$ the reconstruction efficiency and $1+\delta$ the radiative correction factor including initial state radiation and the vacuum polarization. The $\mathcal{B}$ factor is the product of the $\Lambda\to p\pi^-$ and $\bar{\Lambda}\to\bar{p}\pi^+$ branching fractions~\cite{PDG}. From the Born cross section, the effective form factor $G(q^2)$ can be calculated using $G(q^2) = \sqrt{\frac{\sigma_{B\bar{B}}(s)}{(1+\frac{1}{2\tau})(\frac{4\pi\alpha^2\beta }{3s})}}$. In our measurement, the cross section and the effective form factor are found to be $\sigma=119.0\pm5.3(\rm stat.)\pm5.1(\rm sys.)$~pb and $|G|=0.123\pm0.003(\rm stat.)\pm0.003(\rm sys.)$, respectively. The combination of $G$, $R$ and $\Delta\Phi$ enable the first complete calculation of the EMFFs for any baryon.

\end{document}